# Establishment of the conserved operators using variational principle

Nikolay Dementev

*Department of Chemistry, Temple University, Philadelphia, PA 19122*

Operators play a substantial role in mathematical formalism of quantum mechanics. However, explicit forms of the operators are usually postulated, based on the intuitive assumptions. In this study, variational principle was applied to the basic equation for expectation value to vary a generalized form of operator while keeping psi-function invariable. A restriction of being expectation value invariable, allowed one to derive all possible forms of the operators corresponding to the conserved physical entities.

As a result, it was found that only three distinctive forms of the conserved operators are possible, tentatively assigned to be angular momentum-like, momentum-like and total energy-like operators. Surprisingly, all operators included constant, the same one for each of the operators, therefore, making operators the quantum ones. Absence of the quantization in original assumptions suggests that quantum character of the operators and, therefore, the physical entities, is a direct consequence of the existence of the conservation laws.

1. **Introduction**

Formalism of quantum mechanics is hardly imagined without use of operators. Operators are aimed to extract useful information about system by acting on a wave function $\Psi$ the system is represented by. For example, mean value $\overline{L}$ of some physical entity $L$ can be obtained from:

$$\langle \Psi | \hat{L} | \Psi \rangle = \overline{L} \tag{1}$$

where $\hat{L}$ is an operator associated with $L$.[1-4]

Unfortunately, explicit form of an operator, in general, does not coincide with that of a physical entity and, for this reason, is usually postulated.[1-4] In this report, variational principle was applied to eq.(1) to generate all forms, operators of conserved physical entities can possibly have. Variational principle is routinely used in quantum mechanics to obtain an explicit form of $\Psi$ function of the system by varying adjustable parameters in $\Psi$ in such a way as to get $\Psi'$, which corresponds to the minimal energy $E_{min}$ of the system, tentatively assuming that $\Psi'$ will be the best fit for the actual wave function:

$$\langle \Psi | \hat{H} \, \delta ( | \Psi \rangle ) = \delta E \tag{2}$$

$$\Downarrow$$

$$\langle \Psi' | \hat{H} | \Psi' \rangle = E_{min} \tag{3}$$

where $\hat{H}$ is Hamiltonian of the system.[1-4]

One can notice, that the form of operator (i.e. Hamiltonian) in conventional use of the variational principle, is assumed to be known and, thus, invariable. In this study, however, variation of eq.(1) with respect to the form of $\hat{L}$ is explored with assumption of $\Psi$ to be invariable:

$$\langle \Psi | \delta \hat{L} | \Psi \rangle = \delta \overline{L} \tag{4}$$

One gets a useful boundary condition by restricting $\hat{L}$ to be an operator associated with conserved physical entity:

$$\delta \overline{L}_{cons} = 0 \tag{5}$$

Substituting eq.(5) into eq.(4) gives:

$$\langle \Psi | \delta \hat{L}_{cons} | \Psi \rangle = 0 \tag{6}$$

Eq.(6), however, is too general, containing wave vectors, and, for this reason not suiting well for the purpose of finding the forms of $\hat{L}_{cons}$ in a conventional $\Psi$ function variable representation. That is why, prior the calculations, form of a $\Psi$ function with respect to its variables needs to be specified. Assuming periodicity and complexity to be the most essential properties of a $\Psi$ function, one can present $\Psi$ as:

$$\Psi(\varphi) = \rho\, e^{i\varphi} \tag{7}$$

where  $i$ is the imaginary unit;

$\rho$ - amplitude factor;

$\varphi$ - generalized complex variable, which might contain the other variables

(e.g. position $x$, time $t$, etc.)

General form of $\hat{L}_{cons}$ (analogous to any other $\hat{L}$, described elsewhere [4]) can be presented in terms of $\varphi$ as:

$$\hat{L}_{cons} = A_0(\varphi) + A_1(\varphi)\frac{\partial}{\partial \varphi} + A_2(\varphi)\frac{\partial^2}{\partial \varphi^2} + A_3(\varphi)\frac{\partial^3}{\partial \varphi^3} + .... + A_n(\varphi)\frac{\partial^n}{\partial \varphi^n} + .... \quad (8)$$

where $A_0(\varphi), A_1(\varphi), A_2(\varphi), A_3(\varphi),...., A_n(\varphi),...$ are some analytical functions.

The goal of this study, thus, is to find $A_0(\varphi), A_1(\varphi), A_2(\varphi), A_3(\varphi),...., A_n(\varphi),...$ by substituting eq.(7), (8) into:

$$\int \Psi^*(\varphi)\, \delta(\hat{L}_{cons})\, \Psi(\varphi)\, d\varphi = 0 \quad (9)$$

## 2. Derivations

Substitution of (8) into (9) gives:

$$\int \Psi^*(\varphi)\, \delta[A_0(\varphi) + A_1(\varphi)\frac{\partial}{\partial \varphi} + A_2(\varphi)\frac{\partial^2}{\partial \varphi^2} + ...$$
$$... + A_n(\varphi)\frac{\partial^n}{\partial \varphi^n} + ...]\Psi(\varphi)\, d\varphi = 0 \quad (10)$$

Because of the assumption of invariability of $\Psi(\varphi)$, (10) is equivalent to:

$$\int \Psi^*(\varphi)\, \delta\left\{[A_0(\varphi) + A_1(\varphi)\frac{\partial}{\partial \varphi} + A_2(\varphi)\frac{\partial^2}{\partial \varphi^2} + ...\right.$$
$$\left.... + A_n(\varphi)\frac{\partial^n}{\partial \varphi^n} + ....]\Psi(\varphi)\right\} d\varphi = 0 \quad (11)$$

Because of (7) a following relation holds:

$$\frac{\partial^{(n)}}{\partial \varphi^{(n)}} \Psi(\varphi) = i^{(n)} \Psi(\varphi) \qquad (12)$$

$$\Downarrow$$

$$\frac{\partial}{\partial \varphi} \Psi(\varphi) = -\frac{\partial^3}{\partial \varphi^3} \Psi(\varphi) = \frac{\partial^5}{\partial \varphi^5} \Psi(\varphi) = -\frac{\partial^7}{\partial \varphi^7} \Psi(\varphi) = ... \qquad (13)$$

$$\frac{\partial^2}{\partial \varphi^2} \Psi(\varphi) = -\frac{\partial^4}{\partial \varphi^4} \Psi(\varphi) = \frac{\partial^6}{\partial \varphi^6} \Psi(\varphi) = -\frac{\partial^8}{\partial \varphi^8} \Psi(\varphi) = ... \qquad (14)$$

With the help from (13) and (14), (11) can be rearranged as:

$$\int \Psi^*(\varphi) \, \delta \left\{ [A_0(\varphi) + (A_1(\varphi) - A_3(\varphi) + ..)\frac{\partial}{\partial \varphi} + \right.$$

$$\left. + (A_2(\varphi) - A_4(\varphi) + ...)\frac{\partial^2}{\partial \varphi^2}] \Psi(\varphi) \right\} d\varphi = 0 \qquad (15)$$

$$\Downarrow$$

$$\int \Psi^*(\varphi) \, \delta \left\{ [A_0(\varphi) + B_1(\varphi)\frac{\partial}{\partial \varphi} + B_2(\varphi)\frac{\partial^2}{\partial \varphi^2}] \Psi(\varphi) \right\} d\varphi = 0 \qquad (16)$$

where $\qquad B_1(\varphi) = A_1(\varphi) - A_3(\varphi) + A_5(\varphi) - A_7(\varphi) + A_9(\varphi) - ....\qquad (17)$

$$B_2(\varphi) = A_2(\varphi) - A_4(\varphi) + A_6(\varphi) - A_8(\varphi) + A_{10}(\varphi) - .... \qquad (18)$$

Thus, a generalized $\hat{L}_{cons}$ at this step looks like:

$$\hat{L}_{cons} = A_0(\varphi) + B_1(\varphi)\frac{\partial}{\partial \varphi} + B_2(\varphi)\frac{\partial^2}{\partial \varphi^2} \qquad (19)$$

To find the functions $A_1(\varphi), A_1(\varphi), A_2(\varphi), A_3(\varphi), A_4(\varphi),...$ one has to conduct a variation of the following equation, which is equivalent to (16) due to the invariability of $\Psi(\varphi)$:

$$\int \Psi^*(\varphi)\, \delta\left\{ A_0(\varphi) + B_1(\varphi)\frac{\partial}{\partial \varphi} + B_2(\varphi)\frac{\partial^2}{\partial \varphi^2} \right\} \Psi(\varphi)\, d\varphi = 0 \qquad (20)$$

$$\Downarrow$$

$$\int \Psi^*(\varphi) \left\{ \delta A_0(\varphi) + \delta B_1(\varphi)\frac{\partial}{\partial \varphi} + B_1(\varphi)\delta\left(\frac{\partial}{\partial \varphi}\right) + \right.$$

$$\left. + \delta B_2(\varphi)\frac{\partial^2}{\partial \varphi^2} + B_2(\varphi)\delta\left(\frac{\partial^2}{\partial \varphi^2}\right) \right\} \Psi(\varphi)\, d\varphi = 0 \qquad (21)$$

$$\Downarrow$$

$$\int \Psi^*(\varphi) \left\{ \delta A_0(\varphi) + \delta B_1(\varphi)\frac{\partial}{\partial \varphi} + B_1(\varphi)\frac{\partial^2}{\partial \varphi^2} + \right.$$

$$\left. + \delta B_2(\varphi)\frac{\partial^2}{\partial \varphi^2} + B_2(\varphi)\frac{\partial^3}{\partial \varphi^3} \right\} \Psi(\varphi)\, d\varphi = 0 \qquad (22)$$

$$\Downarrow (12)$$

$$\int \Psi^*(\varphi)\, \{\delta A_0(\varphi) + i\delta B_1(\varphi) - B_1(\varphi) - \delta B_2(\varphi) - i B_2(\varphi)\}\, \Psi(\varphi)\, d\varphi = 0 \qquad (23)$$

$$\Downarrow$$

$$\int \Psi^*(\varphi)\, \{[\delta A_0(\varphi) + i\delta B_1(\varphi) - \delta B_2(\varphi)] - [B_1(\varphi) + i B_2(\varphi)]\}\, \Psi(\varphi)\, d\varphi = 0 \qquad (24)$$

$$\Downarrow$$

$$\int \Psi^*(\varphi)\, \{\delta[A_0(\varphi) + i B_1(\varphi) - B_2(\varphi)] - [B_1(\varphi) + i B_2(\varphi)]\}\, \Psi(\varphi)\, d\varphi = 0 \qquad (25)$$

The latter equation has to be valid at any particular value of an increment $\delta[A_0(\varphi) + iB_1(\varphi) - B_2(\varphi)]$, therefore, (25) should also be valid in the limit case of $\delta[A_0(\varphi) + iB_1(\varphi) - B_2(\varphi)] \to 0$. For this reason, (25) can be simplified to:

$$\int \Psi^*(\varphi)[B_1(\varphi) + iB_2(\varphi)] \Psi(\varphi)\, d\varphi = 0 \tag{26}$$

Since $\Psi^*(\varphi)\Psi(\varphi) \neq 0$ the above equation can be true only if:

$$B_1(\varphi) = -iB_2(\varphi) \tag{27}$$

Substitution of (27) into (24) gives:

$$\int \Psi^*(\varphi)[\delta A_0(\varphi) + i\delta B_1(\varphi) - \delta B_2(\varphi)]\Psi(\varphi)\, d\varphi = 0 \tag{28}$$

$$\Downarrow$$

$$\delta A_0(\varphi) + i\delta B_1(\varphi) - \delta B_2(\varphi) = 0 \tag{29}$$

Substitution of (27) into (29) gives:

$$\delta A_0(\varphi) + i\delta[-iB_2(\varphi)] - \delta B_2(\varphi) = 0 \tag{30}$$

$$\Downarrow$$

$$\delta A_0(\varphi) + \delta B_2(\varphi) - \delta B_2(\varphi) = 0 \tag{31}$$

$$\Downarrow$$

$$\delta A_0(\varphi) = 0 \tag{32}$$

$$\Downarrow$$

$$A_0(\varphi) = A = const \tag{33}$$

Substitution of (27) and (33) into (19) gives:

$$^0\hat{L}_{cons} = A - iB_2(\varphi)\frac{\partial}{\partial\varphi} + B_2(\varphi)\frac{\partial^2}{\partial\varphi^2} \tag{34}$$

Upon the substitution of (34) into:

$$\overline{L} = \int \Psi^*(\varphi)\,\hat{L}_{cons}\,\Psi(\varphi)\,d\varphi = 0 \tag{35}$$

one gets:

$$\overline{L} = \int \Psi^*(\varphi)\left[A - iB_2(\varphi)\frac{\partial}{\partial\varphi} + B_2(\varphi)\frac{\partial^2}{\partial\varphi^2}\right]\Psi(\varphi)\,d\varphi = 0 \tag{36}$$

$$\Downarrow (12)$$

$$\overline{L} = \int \Psi^*(\varphi)[A]\Psi(\varphi)\,d\varphi = 0 \tag{37}$$

which implies that:

$$\hat{L}_{cons} = A = const \tag{38}$$

if $\hat{L}_{cons}$ of form (34) is used.

There are six special cases of (34):

$$^1\hat{L}_{cons} = A \tag{39}$$

$$^2\hat{L}_{cons} = -iB_2(\varphi)\frac{\partial}{\partial\varphi} \tag{40}$$

$$^3\hat{L}_{cons} = B_2(\varphi)\frac{\partial^2}{\partial\varphi^2} \tag{41}$$

$$^4\hat{L}_{cons} = A - iB_2(\varphi)\frac{\partial}{\partial\varphi} \tag{42}$$

$$^5\hat{L}_{cons} = A + B_2(\varphi)\frac{\partial^2}{\partial\varphi^2} \tag{43}$$

$$^6\hat{L}_{cons} = -iB_2(\varphi)\frac{\partial}{\partial\varphi} + B_2(\varphi)\frac{\partial^2}{\partial\varphi^2} \qquad (44)$$

Since (35) is valid for both, $^0\hat{L}_{cons}$ and $^k\hat{L}_{cons}$ (where k = 1,2,3,4,5,6), a following equation has to be valid as well:

$$\int \Psi^*(\varphi)[^0\hat{L}_{cons} - {}^k\hat{L}_{cons}]\Psi(\varphi)\,d\varphi = 0 \qquad (45)$$

By solving (i.e. finding $B_2(\varphi)$) each of the special cases, one can find that only three distinctive forms of $\hat{L}_{cons}$ are possible:

$$\hat{\alpha}_\varphi = A \qquad \hat{\beta}_\varphi = -iA\frac{\partial}{\partial\varphi} \qquad \hat{\gamma}_\varphi = A\frac{\partial^2}{\partial\varphi^2} \qquad (46)$$

### 3. Summary and Concluding Remarks

Even though conservation operators (46) in this work were obtained in their general form (i.e. in the generalized variable $\varphi$ representation), some conclusions are still can be made:

1. Only three distinctive operators of the conserved variables are possible;
2. In the case when:

$$\varphi = x \qquad (47)$$

where $x$ is a position variable, operators (46) transform into:

$$\hat{\alpha}_x = A \qquad \hat{\beta}_x = -iA\frac{\partial}{\partial x} \qquad \hat{\gamma}_x = A\frac{\partial^2}{\partial x^2} \qquad (48)$$

One can easily see, that:

$$\hat{\beta}_x = \hat{p}_x = -i\hbar \frac{\partial}{\partial x} \qquad (49)$$

if
$$A = \hbar \qquad (50)$$

where $\quad \hat{p}_x$ is a momentum operator;

$\hbar$ is the Planck's constant

In the assumption of (50), operators (48) can be rewritten as:

$$\hat{\alpha}_x = \hbar \qquad \hat{\beta}_x = -i\hbar \frac{\partial}{\partial x} \qquad \hat{\gamma}_x = \hbar \frac{\partial^2}{\partial x^2} \qquad (51)$$

(51) allows one to interpret $\hat{\alpha}_x$ and $\hat{\beta}_x$ as operators of angular momentum and momentum, respectively.[1-4] The only option remaining is to interpret $\hat{\gamma}_x$ as an operator of the total energy. From this point of view, operators $\hat{\alpha}_\varphi$, $\hat{\beta}_\varphi$ and $\hat{\gamma}_\varphi$ can be called as angular momentum-like, momentum-like, and the total energy-like operators, respectively.[5]

3. Operators of the conserved quantities have to be quantum operators. Indeed, each of the operators (46) contains $A = const$ (i.e. the operators are the quantum ones) and this became possible due to the restriction (5) which is equivalent to the conservation requirement.